\documentclass[12pt]{article}
\usepackage{graphicx}

\newcommand\ion[2]{#1$\;${\small\uppercase\expandafter{\romannumeral #2\relax}}}
\newcommand{\afx}{PS1-10afx}

\newcommand{\zps}{\ensuremath{z_{\rm P1}}}

\newcommand{\msun}{${\rm M}_\odot$}          
\newcommand{\simlt}{\mathrel{\hbox{\rlap{\hbox{\lower4pt\hbox{$\sim$}}}\hbox{$<$}}}}
\newcommand{\simgt}{\mathrel{\hbox{\rlap{\hbox{\lower4pt\hbox{$\sim$}}}\hbox{$>$}}}}

\newcommand{\kps}{km s$^{-1}$}

\newcommand{\eps}{erg\,s$^{-1}$}
\newcommand{\SNIa}{SNIa}

\newcommand{\arcsec}{$^{\prime\prime}$}

\newcommand{\mnras}{MNRAS}
\newcommand{\apj}{ApJ}
\newcommand{\apjl}{ApJ}
\newcommand{\nat}{Nature}

\newcommand{\aap}{A\&A}

\newcommand{\pasp}{PASP} 
\newcommand{\aj}{AJ}
\newcommand{\pasj}{PASJ} 

\usepackage{scicite}
\usepackage{times}

\newenvironment{sciabstract}{%
\begin{quote} \bf}
{\end{quote}}

\newcounter{lastnote}
\newenvironment{scilastnote}{%
\setcounter{lastnote}{\value{enumiv}}%
\addtocounter{lastnote}{+1}%
\begin{list}%
{\arabic{lastnote}.}
{\setlength{\leftmargin}{.22in}}
{\setlength{\labelsep}{.5em}}}
{\end{list}}

\title{Detection of the Gravitational Lens Magnifying a Type Ia Supernova} 

\author
{Robert M. Quimby,$^{1\ast}$
  Masamune Oguri,$^{1,2}$
  Anupreeta More,$^{1}$
  Surhud More,$^{1}$\\
  Takashi J. Moriya,$^{3,4}$
  Marcus C. Werner,$^{1}$
  Masayuki Tanaka,$^{5}$
  Gaston Folatelli,$^{1}$\\
  Melina C. Bersten,$^{1}$
  Keiichi Maeda,$^{6}$ and
  Ken'ichi Nomoto,$^{1}$\\
\\
\normalsize{$^{1}$Kavli Institute for the Physics and Mathematics of the Universe (WPI),}\\
\normalsize{Todai Institutes for Advanced Study, The University of Tokyo,}\\
\normalsize{5-1-5 Kashiwanoha, Kashiwa-shi, Chiba, 277-8583, Japan}\\
\normalsize{$^{2}$Department of Physics, The University of Tokyo, Tokyo 113-0033, Japan}\\
\normalsize{$^{3}$Argelander Institute for Astronomy, University of Bonn}\\
\normalsize{Auf dem H\"ugel 71, D-53121 Bonn, Germany}\\
\normalsize{$^{4}$ Research Center for the Early Universe, Graduate School of Science}\\
\normalsize{University of Tokyo, Hongo 7-3-1, Bunkyo, Tokyo 113-0033, Japan}\\
\normalsize{$^{5}$ National Astronomical Observatory of Japan}\\
\normalsize{2-21-1 Osawa, Mitaka, Tokyo 181-8588, JAPAN}\\
\normalsize{$^{6}$ Department of Astronomy, Kyoto University, Kitashirakawa-Oiwake-cho}\\
\normalsize{Sakyo-ku, Kyoto 606-8502, Japan}\\
\\
\normalsize{$^\ast$To whom correspondence should be addressed; E-mail:  robert.quimby@ipmu.jp.}
}

\date{}

\begin{document} 
\baselineskip24pt
\maketitle 

\begin{sciabstract}
Objects of known brightness, like Type Ia supernovae (\SNIa), can be
used to measure distances. If a massive object warps spacetime to form
multiple images of a background \SNIa, a direct test of cosmic
expansion is also possible. However, these lensing events must first
be distinguished from other rare phenomena. Recently, a supernova was
found to shine much brighter than normal for its distance, which
resulted in a debate: was it a new type of superluminous supernova or
a normal SNIa magnified by a hidden gravitational lens? Here we report
that a spectrum obtained after the supernova faded away shows the
presence of a foreground galaxy--the first found to strongly magnify a
\SNIa. We discuss how more lensed \SNIa\ may be found than previously
predicted.
\end{sciabstract}

A peculiar supernova, \afx, was discovered by the Panoramic Survey
Telescope \& Rapid Response System 1 (Pan-STARRS1) on 2010 August 31
(UT) \cite{chornock2013}. The unusually red color of the object
spurred the Pan-STARRS1 team to conduct an array of follow-up
observations including optical and near infrared spectroscopy, which
yielded a redshift of $z=1.39$. Combined with relatively bright
photometric detections, this redshift would imply a peak luminosity of
$4 \times 10^{44}$\,\eps, which is 400 times brighter than the typical
core-collapse supernova. A rare class of superluminous supernovae
(SLSN) \cite{galyam2012} have shown similarly high bolometric
outputs, but \afx\ distinguishes itself from all other SLSN on two
important counts: \afx\ is much redder (cooler) and evolved much
faster than any SLSN. A generic feature of SLSN models
\cite{rakavy1967,barkat1967,woosley2007,woosley2010,kasen_bildsten2010,chevalier_irwin2011,moriya2013}
is that they employ high temperatures and/or large photospheric radii
to generate high luminosities (recalling that $L \propto T^4R^2$). The
observations of \afx\ do not fit with these models, suggesting that if
it is a SLSN, it is in a class of its own.

An alternate hypothesis \cite{quimby2013b} is that \afx\ is actually a
regular Type Ia supernova (\SNIa) with a normal luminosity, but its
apparent brightness has been magnified by a gravitational
lens. Spectra of \afx\ are well fit by normal \SNIa\ templates, as are
the colors and light curve shapes. However, normal \SNIa\ exhibit a
tight relation between the widths of their light curves and their peak
luminosities \cite{phillips1999,jha2007,hicken2009,sullivan2010}, and
\afx\ appears 30 times brighter than expected according to this
relation. Such a large magnification of brightness can only occur
naturally from strong gravitational lensing, whereby the light
emanating from the supernova is bent to form an Einstein-Chwolson
ring, or several discrete magnified images (typically two or four) if
the alignment is not axisymmetric. Pan-STARRS1 has surveyed sufficient
volume to expect such a chance alignment \cite{oguri_marshall2010,som}
and it is possible that the angular extent of the lensed images was
simply too small to be resolved by the observations
available. However, for this hypothesis to be confirmed, we must
explain why the existing observations give such conclusive photometric
and spectroscopic evidence for the presence of the supernova's host
galaxy, but the same observations fail to obviously indicate the
presence of a foreground lens.

\begin{figure}
\begin{center}
 \includegraphics[width=\linewidth]{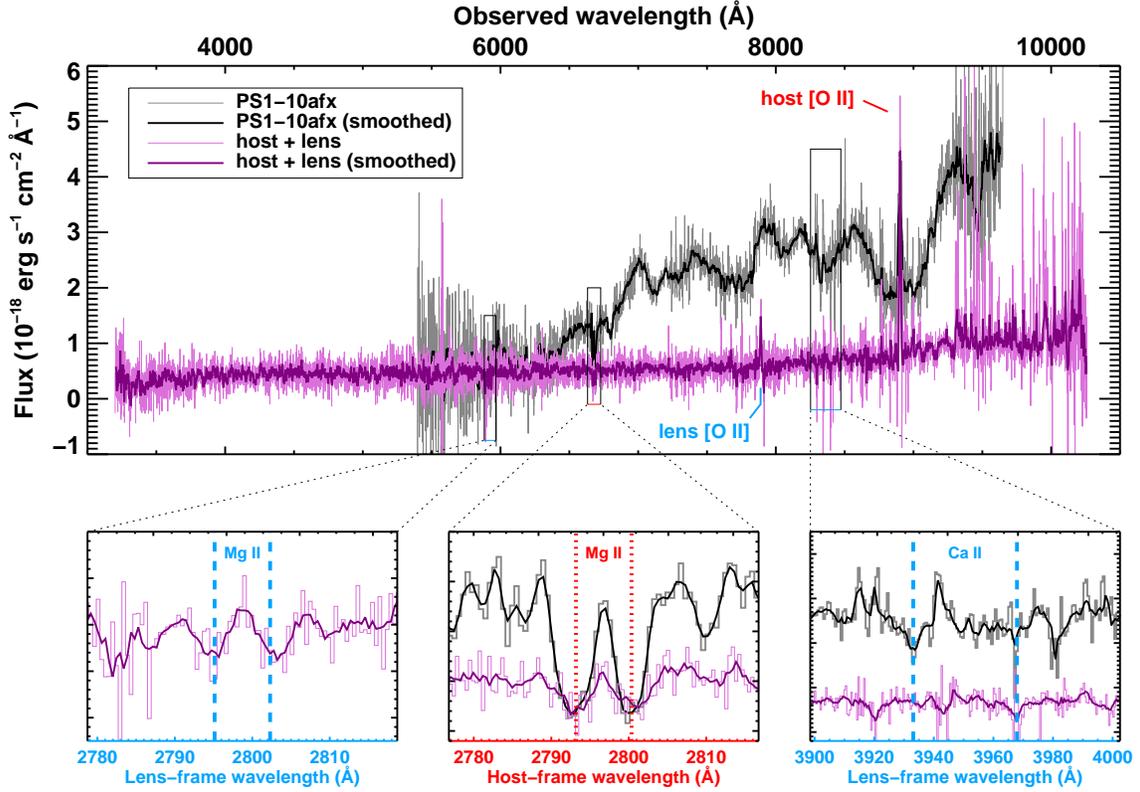} 
 \caption{ {\bf Spectra of the quiescent light at the location of
     \afx.} The Keck/LRIS observations (purple) taken about 450 rest
   frame days after the supernova reached maximum light show the
   presence of two emission features, which we identified as
   [\ion{O}{2}] from galaxies at $z=1.1168$ and $z=1.3885$. We also
   detected absorption lines corresponding the \ion{Mg}{2} doublet for
   both the foreground lens galaxy (lower left panel; blue labels) and
   the host (lower middle panel; red labels). Vertical lines mark the
   rest frame wavelengths for the doublets in the lower panels. A
   spectrum of \afx\ taken near maximum light \cite{chornock2013} is
   shown in gray for comparison. This supernova spectrum has been
   shifted slightly to align the \ion{Mg}{2} features with the
   Keck/LRIS data \cite{som}. The lower-right panel shows that the
   supernova light may be absorbed at wavelengths corresponding to
   \ion{Ca}{2} H\&K in the foreground galaxy. \ion{Ca}{2} H is
   coincident with a strong night sky line, but the \ion{Ca}{2} K
   falls in a relatively clean spectral region. }
   \label{fig:compare_spec}
\end{center}
\end{figure}

We used the Keck-I telescope with the Low-Resolution Imaging
Spectrograph (LRIS) \cite{oke1995} with the upgraded red channel
\cite{rockosi2010} to observe the host galaxy and any foreground
objects at the sky position of \afx\ on 2013 September 7 (see
Fig. \ref{fig:slit}) \cite{som}. As illustrated in figure
\ref{fig:compare_spec}, there are two narrow emission features that
persist at the location of \afx\ now that the supernova itself has
faded away. The [\ion{O}{2}] emission doublet ($\lambda\lambda =
3726.1,3728.8$\,\AA\ in the rest frame) from the host galaxy
previously identified \cite{chornock2013} is clearly recovered
(Fig. \ref{fig:host_OII}), but we additionally detected a second
emission line at about 7890\,\AA. Because there are no strong emission
lines expected from the host at this wavelength ($\sim$3300\,\AA\ in
the host frame), this detection suggests the presence of a second
object coincident with \afx.

The most probable identification for the 7890\,\AA\ feature is
[\ion{O}{2}] at $z=1.1168 \pm 0.0001$. At this redshift, other strong
emission lines such as H-beta or [\ion{O}{3}] would lie outside of our
wavelength coverage. However, as depicted in figure
\ref{fig:compare_spec}, we detected a \ion{Mg}{2} absorption doublet
($\lambda\lambda = 2795.5,2802.7$\,\AA\ in the rest frame) at
$z=1.1165 \pm 0.0001$. Blueshifted absorption outflows are typical of
star-forming galaxies \cite{erb2012}, so this estimate is compatible
with that derived from the emission lines. We also identify possible
\ion{Mg}{1} ($\lambda = 2853.0$) and \ion{Fe}{2} ($\lambda = 2344.2,
2373.7, 2382.8, 2586.7, 2600.2$) lines with a matching redshift. Given
these identifications and the extended nature of the emission (see
Fig. \ref{fig:lens_OII}), it is clear that the second object is a
galaxy lying in front of \afx\ and its host. Indeed, the near maximum
light spectra of \afx\ \cite{chornock2013} show what could be
\ion{Ca}{2} absorption from this foreground galaxy
(Fig. \ref{fig:compare_spec}).

\begin{figure}
\begin{center}
 \includegraphics[width=\linewidth]{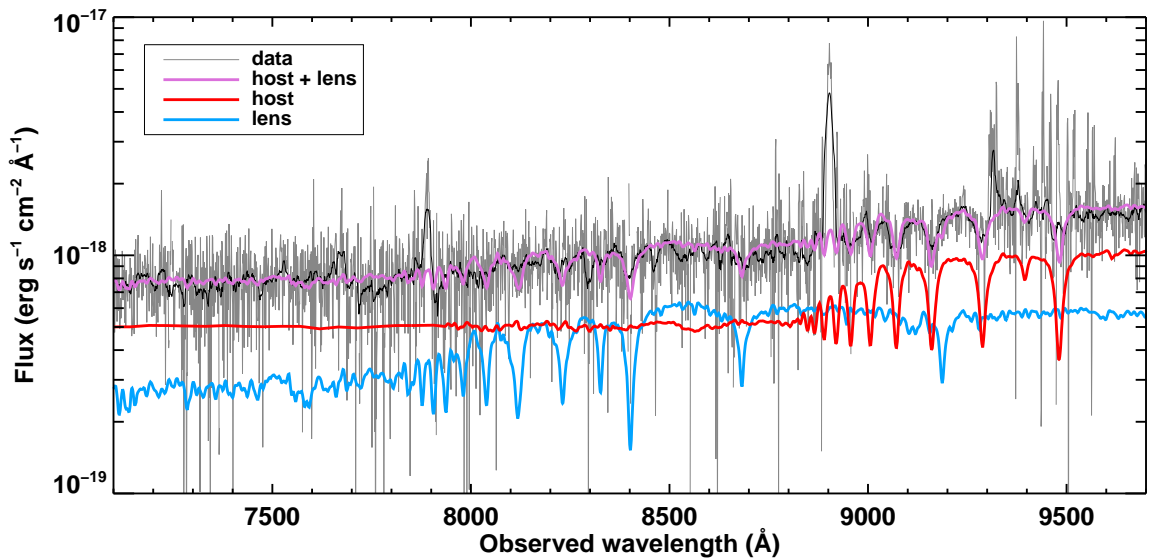} 
 \caption{ {\bf Decomposition of the observed spectra into lens and
     host galaxy components.} We modeled the lens (blue line) and host
   (red line) as single stellar populations at $z=1.1168$ and
   $z=1.3885$, respectively. We varied the age and total stellar mass
   of each galaxy in order to find the sum (purple line) that best
   matched the observed spectra (gray; smoothed spectra in black). The
   models only include starlight; light emitted by gas, such as the
   [\ion{O}{2}] lines seen at about 8900\,\AA\ and 7900\,\AA, is
   neglected. }
   \label{fig:ssp}
\end{center}
\end{figure}

With its redshift secure, we next checked whether the foreground
galaxy can satisfy the lens requirements of \afx
\cite{quimby2013b}. To do this, we derived a stellar mass for the
foreground galaxy and used an empirical relation between stellar mass
and 1-D velocity dispersion as explained below. We fitted a set of
single stellar population (SSP) models to the combined spectra to
measure the stellar masses of the host and foreground galaxies (see
Figs. \ref{fig:ssp} and \ref{fig:ssp_lens}) \cite{som}. The best-fit
SSP combination was a $\sim$1\,Gyr old foreground galaxy with a
stellar mass of $(9 \pm 2) \times 10^{9}$\,\msun\ and a more distant
host galaxy with $(7 \pm 1) \times 10^{9}$\,\msun\ and a younger
($\sim$0.1\,Gyr) population. The extinction in the foreground galaxy
is consistent with zero ($A_V = 0.28^{+0.48}_{-0.28}$), but the host
galaxy requires significant reddening ($A_V = 1.62 \pm 0.18$).

\begin{figure}
\begin{center}
 \includegraphics[width=\linewidth]{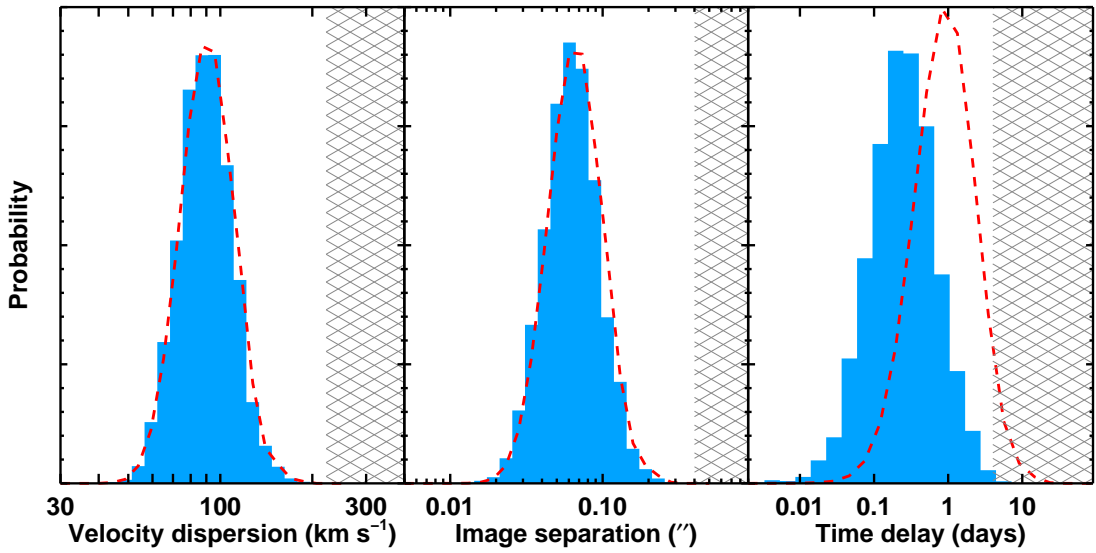}
 \caption{ {\bf Probability distributions for the lens parameters.}
   The panels show, left to right, the relative likelihood for the
   line of sight velocity dispersion of the lensing galaxy, the
   maximum separation between lensed images, and the maximum time
   delay between lensed images as predicted by our Monte Carlo
   simulation. The blue histograms account for the total
   magnification, $\mu = 31 \pm 5$, measured for \afx
   \cite{quimby2013b}, and the dashed red curves neglect this prior.
   The hatched areas are excluded based on the observations of \afx,
   specifically the lack of resolved images or evidence for time
   delays.  }
   \label{fig:lens}
\end{center}
\end{figure}

Stellar mass contributes only a fraction of a galaxy's total mass,
which is usually dominated by dark matter. The ratio of these masses
varies from galaxy to galaxy, but they are strongly correlated. Using
the stellar-mass-to-velocity-dispersion relation measured from the
SDSS DR7 spectroscopic sample \cite{kauffmann2003,blanton2005} (see
Fig. \ref{fig:mstar-sigma}), we inferred a probability distribution
for the foreground galaxy's velocity dispersion \cite{som}.  We then
used this as input to a Monte Carlo simulation from which we derived
the posterior distributions for the lens parameters
(Fig. \ref{fig:lens}). We find that the redshift and mass of the
foreground galaxy make it fully consistent with a gravitational lens
that is capable of satisfying the magnification, image separation
limits, and time delay constraints of \afx\ \cite{quimby2013b}. We
thus concluded that \afx\ is not a superluminous supernova but a
normal \SNIa\ magnified by a strong gravitational lens at $z=1.1168$.

Our new data further explain why the lensing galaxy was not evident in
prior observations.  Even though the host galaxy is slightly less
massive, more extinguished, and farther away than the lensing galaxy,
it harbors a younger stellar population that shines more brightly per
unit mass. Because of this, the foreground object is only comparable
in brightness to the host over a narrow range of wavelengths longer
than the lensing galaxy's 4000\,\AA\ break but shorter than the
host's. This makes it difficult to see the light from the lens galaxy
over the glare from the host galaxy.

The lack of multiple images or signs of time delay from \afx\ can also
be explained from our Monte Carlo simulations. The high total
magnification of \afx\ is best recovered from alignments that produce
four images (79\% probability of a ``quad'' system), but the maximum
separation between the different multiple images is small ($\Delta
\theta < 0.12$\arcsec\ at 95\% confidence) and the maximum phase delay
is short ($\Delta t < 1.3$\,day at 95\% confidence). The available
observations of \afx\ are thus likely insufficient to resolve any
effects from the gravitational lensing other than its magnification.

In the future, high angular resolution imaging enabled by adaptive
optics (AO) or space-based resources like the Hubble Space Telescope
({\it HST}) could be used to spatially resolve the multiple images of
gravitational lensed \SNIa\ like \afx. This would not only provide
immediate confirmation that gravitational lensing is at play, but it
would also provide important constraints on the nature of the lens. In
theory, multiple epochs of high resolution imaging could be used to
measure the time delay between each image and the magnification
factors for each. Such observations could yield strong constraints on
cosmic expansion.

Because the universe is expanding, the path lengths of the more
delayed images will be stretched more by cosmic expansion, and the
magnitude of this delay is directly tied to the Hubble parameter
\cite{refsdal1964}. However, the time delay is also dependent on the
mass density profile of the lens, which will be dictated by
unobservable dark matter. For objects of known brightness, like
\SNIa, we can use the readily measurable magnification to break this
degeneracy \cite{oguri_kawano2003}. As is likely the case for \afx,
spectra taken near maximum light can reveal not only the redshift of
the supernova but, using absorption line spectroscopy, the redshift
and velocity dispersion of the lens as well. (Note that at maximum
light, the resolving power available from Keck/LRIS may have been
sufficient to resolve the \ion{Ca}{2} K line from the lens, but the
signal-to-noise ratio and resolution of the available observations is
too low: $\sigma < 125$\,\kps\ at 95\% confidence.) Thus, future
discoveries of gravitationally lensed \SNIa\ may be used to make a
direct and precise measurement of the Hubble constant, but only if the
needed follow-up observations commence in a timely manner.

\begin{figure}
\begin{center}
 \includegraphics[width=\linewidth]{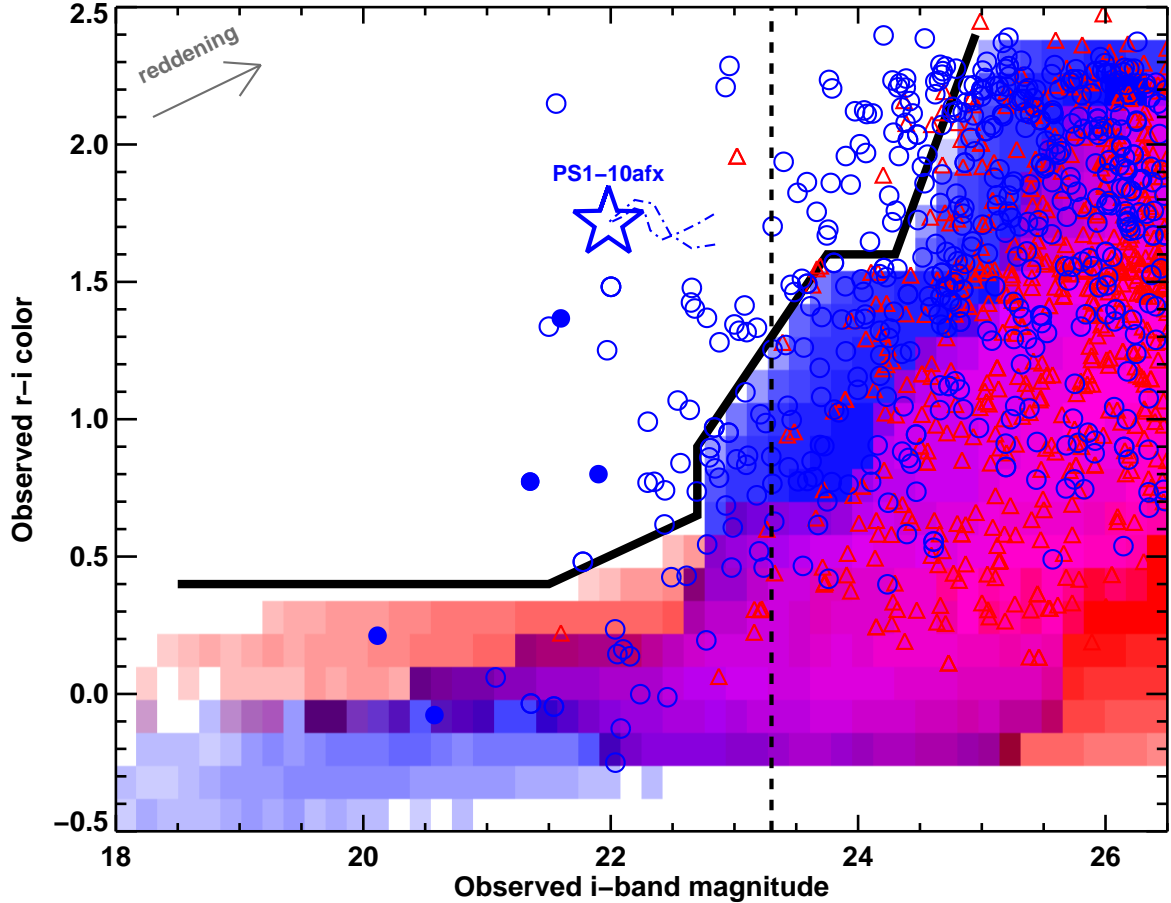} 
 \caption{ {\bf Color-magnitude diagram showing how lensed \SNIa\ can
     be distinguished from un-lensed events.} The blue shaded area
   shows the expected color-magnitude distribution for un-lensed
   \SNIa\ on a log scale, and the red shaded area corresponds to
   core-collapse supernovae. The $r-i$ colors for low redshift
   supernovae are relatively blue. However, at higher redshifts
   (fainter observed magnitudes), the color becomes red as the peak of
   the rest-frame spectral energy distribution passes through the
   observer-frame bands. The red limit for un-lensed supernovae at a
   given $i$-band magnitude is denoted by the thick black line. Blue
   circles and red triangles show the distribution of lensed
   \SNIa\ and core-collapse supernovae, respectively, predicted by
   Monte Carlo simulations \cite{som}. Filled symbols indicate objects
   that could be resolved from ground based observations, such as
   those planned by the Large Synoptic Survey Telescope (LSST). Open
   symbols depict objects that require high angular resolution
   follow-up observations to resolve spatially. The open star marks
   the values corresponding to the peak $i$-band brightness of \afx,
   and the dash-dotted curve shows that the color evolution within one
   magnitude of this peak is minimal. The vertical dashed line marks
   the single epoch limit predicted for LSST. The arrow shows the
   reddening vector, assuming $A_V=1.0$\,mag. }
   \label{fig:cmd}
\end{center}
\end{figure}

To begin follow-up observations of lensed \SNIa\ candidates while they
are still on the rise (needed for accurate delay time measurements) or
near maximum light (the optimal phase for absorption line
spectroscopy), an efficient vetting process must be employed to
eliminate the non-lensed supernovae that out number lensed \SNIa\ by a
few thousand to one. A means to accomplish this feat was demonstrated
by the selection of \afx\ for follow-up by the Pan-STARRS1 team --
they were first motivated to examine \afx\ based on its unusually red
color \cite{chornock2013}. In figure \ref{fig:cmd} we show a
color-magnitude diagram for observed supernovae
\cite{som}. \afx\ showed a color near maximum light of $r-i \sim 1.7$
whereas un-lensed supernovae brighter than the Pan-STARRS1 detection
limits have $r-i < 0.5$. (Note that quiescent galaxy light is removed
from these measurements using pre-supernova images). Since the lensing
probability increases with redshift and, for a given sky area, the
number of supernovae also increases with redshift (because of both
rates and increased volume), the majority of lensed \SNIa\ expected
from a flux limited survey will be high redshift events that will
typically appear much redder than the more nearby population of
un-lensed supernovae. We thus propose that selecting supernovae with
colors redder that the bold line in figure \ref{fig:cmd} (for a given
i-band magnitude) during their rise to peak may be an effective way to
identify lensed supernovae. This will bypass the need to detect
multiple, resolved images and will thus increase the expected number
of lensed SN Ia from future surveys by a factor of 5 over traditional
selection techniques.

The large magnification and relatively low-mass lens galaxy of
\afx\ may prove typical of gravitationally lensed \SNIa\ that will be
discovered by future, flux-limited surveys given selection bias:
brighter objects are easier to detect, and unresolved images formed by
a low-mas lens effectively make a single, brighter object \cite{som}
(see Fig. S6). Alternatively, the lens mass and high magnification of
PS1-10afx may indicate a problem with our understanding of the
starlight to dark matter connection in the early universe. Further
studies of this system are thus warranted. A multi-wavelength,
high-angular-resolution study of the lens and host galaxies with {\it
  HST}, ground based AO, and the Atacama Large
Millimeter/submillimeter Array (ALMA) could further constrain the
lensing model and provide an important reference for future studies of
gravitationally lensed \SNIa.


\begin{scilastnote}
\item {\bf Acknowledgments:} This work was supported in part by the
  Kakenhi Grant-in-Aid for Young Scientists (B)(24740118) from the
  Japan Society for the Promotion of Science, the World Premier
  International Research Center Initiative, MEXT, Japan, the FIRST
  program, ``Subaru Measurements of Images and Redshifts (SuMIRe),''
  and by the Japan Society for the Promotion of Science Research
  Fellowship for Young Scientists (23-5929). The data presented herein
  were obtained at the W.M. Keck Observatory, which is operated as a
  scientific partnership among the California Institute of Technology,
  the University of California and the National Aeronautics and Space
  Administration. The Observatory was made possible by the generous
  financial support of the W.M. Keck Foundation. The data presented in
  this paper are available from the Weizmann Interactive Supernova
  Data Repository (www.weizmann.ac.il/astrophysics/wiserep).
\end{scilastnote}

\clearpage


%
\clearpage

\begin{center}
{\LARGE Supplementary Materials for}

\vspace{0.5cm}
{\bf \large Detection of the Gravitational Lens Magnifying a Type Ia Supernova} 

\hfill
\begin{minipage}{\textwidth}
\begin{center}
\vspace{0.5cm}

Robert M. Quimby,$^{\ast}$
  Masamune Oguri,
  Anupreeta More,
  Surhud More,\\
  Takashi J. Moriya,
  Marcus C. Werner,
  Masayuki Tanaka,
  Gaston Folatelli,\\
  Melina C. Bersten,
  Keiichi Maeda,
  Ken'ichi Nomoto\\
\end{center}
\end{minipage}

\vspace{0.5cm}
\normalsize{$^\ast$To whom correspondence should be addressed; E-mail:  robert.quimby@ipmu.jp.}
\end{center}

\vspace{2cm}
{\bf This PDF file includes:}

\hfill
\begin{minipage}{\dimexpr\textwidth-2.0cm}
\vspace{0.5cm}
Supplementary Text\\
Figures S1 to S6\\
References\\
\end{minipage}

\clearpage
{\bf \large Supplementary Text}

\vspace{10mm}
\underline{Spectroscopy}
\vspace{5mm}

The 2013 September 7 observations from the Keck-I telescope with the
Low-Resolution Imaging Spectrograph (LRIS) \cite{oke1995} employed an
atmospheric dispersion corrector to prevent wavelength dependent
losses for the non-parallactic slit angles required
\cite{filippenko1982}. The $1.0$\arcsec\ slit was oriented to include
a nearby star (see Fig. \ref{fig:slit}), and the telescope was
nodded $\pm2$\arcsec\ along the slit in an ABBA pattern in several
sets over the observation. Light was sorted into red and blue channels
using the 560 dichroic. We used the 400/3400 grism for the blue
channel, giving a spectral resolution, $R = \lambda / \lambda_{\rm
  FWHM}$, of about $500-700$ from 3200\,\AA\ to 5600\,\AA\ as measured
from arc lamp lines. On the red side, we used the 400/8500 grating to
give a spectral resolution of about $1000-1700$ between 5600\,\AA\ and
10300\,\AA. The resolution was determined by noting the FWHM of the
Gaussian kernel that can best convolve a high resolution night sky
spectrum \cite{hanuschik2003} to match the sky lines in our data. We
used spatial binning on the red channel to give a
$0.27$\arcsec\,pixel$^{-1}$ scale, but left the blue side un-binned
($0.135$\arcsec\,pixel$^{-1}$). Integration times were set to 847\,s
in the blue channel and 817\,s in the red channel (per exposure) to
accommodate differences in readout speed. A total of 28 exposures
 (about 6.5\,hours on target) were obtained in each channel under good
sky conditions and $\sim0.7$\arcsec\ seeing.

Two of the blue spectra were found to be anomalous; although the sky
counts are similar to the other exposures in the blue, there is a
pivot point at longer wavelengths beyond which the counts taper off
significantly. These exposures were excluded from our analysis. No
problems were found with the associated red channel exposures,
suggesting the problem is internal to the blue side of the instrument.

At the end of our first 4 ABBA sets on \afx, we imaged the slit and
found that the reference star (and thus the position of \afx) was
slightly offset (about 0.4\arcsec\ West) from the center of the
slit. We imaged the slit prior to and after the final 3 ABBA sets to
ensure and verify that the slit was properly aligned. The first 4 and
second 3 sets are offset along the slit by about $2$\arcsec, thus
placing the target at 4 distinct positions in the detector plane.

We extracted the spectra using IRAF \cite{iraf}, called through python
scripts using the PyRAF package, for basic de-trending and custom IDL
scripts for the final target extraction. We observed the standard
star, BD+28d4211, before, in the middle of, and after the
\afx\ observation for use in spectrophotometric calibration and as a
reference for removing telluric features. To correct for atmospheric
extinction, we used the Mauna Kea extinction curve
\cite{buton2013}. LA-Cosmic \cite{van-dokkum2001} was used to mark
pixels affected by cosmic rays and particle events on individual
frames. We have implemented a 2-D sky subtraction procedure
\cite{kelson2003}. Briefly, the sky is fit with B-splines to the two
dimensional spectra, with masking and iteration used to exclude object
or otherwise deviant pixels. We reduce the data to a 1-D spectrum
using optimal extraction applied to the full data set, without ever
warping the 2-D spectra. For some of our analysis, we directly use the
set of 2-D spectra. The wavelength scale was calibrated first using
afternoon arc lamp exposures, and then adjusted by cross-correlating
to the UVES night sky spectra \cite{hanuschik2003}. By
cross-correlating sections of our final sky spectrum against the UVES
templates, we estimate the corrected wavelengths are accurate to
0.4\,\AA\ in the blue channel ($\lambda < 5600$) and accurate to
0.2\,\AA\ in the red channel. Finally, we apply a small
($-5$\,km\,s$^{-1}$) correction to place the measured wavelengths
into the Heliocentric rest frame (this frame differs from the CMB rest
frame by about $-350$\,km\,s$^{-1}$).

Figures \ref{fig:host_OII} and \ref{fig:lens_OII} show the combined
2-D data from the second (properly aligned) pointing near the location
of the two emission lines. The spatial coordinate has been set by
assuming a separation of 19.13\arcsec\ between \afx\ and the reference
star in the slit (see Fig. \ref{fig:slit}). This coordinate system is
only approximate given the uncertainties in the centroid of
\afx\ ($\sim$0.1\arcsec) \cite{chornock2013}, the systematic offset to
the CFHT coordinate system (assumed to be $\sim$0.1\arcsec), and
non-linearities in the Keck/LRIS detector plane
(est. $\sim$0.1\arcsec). The spatial location of the host, delineated
by the extent of the emission line in figure \ref{fig:host_OII}, is
thus consistent if not slightly south of \afx. Similarly, the spatial
extent of [\ion{O}{2}] from the foreground object
(Fig. \ref{fig:lens_OII}) overlaps the location of \afx\ and appears
to have its centroid some 0.5\arcsec\ north of the host's
[\ion{O}{2}], which further suggests these represent physically
distinct sources.

For the host galaxy, we derive a redshift of $z=1.3885 \pm 0.0001$
from the [\ion{O}{2}] doublet by fitting a 2-D model to the set of
individual exposures. We note that this redshift may be slightly
higher than previously reported \cite{chornock2013}. This systematic
difference could be caused by calibration differences (e.g. sky lines
in the supernova spectra appear systematically offset in wavelength by
about 1.6\,\AA\ compared to the UVES night sky atlas). Figure
\ref{fig:host_OII} suggests a second possibility.  The host's
    [\ion{O}{2}] doublet appears to shift in velocity with spatial
    position. In other words, the galaxy may be rotating and different
    slit orientations may thus lead to slightly biased redshift
    measurements (we similarly find a systematic offset between our
    measurements of the \ion{Mg}{2} absorption line wavelengths and
    those previously reported).

By using the 2-D data to fit for the emission line, we can account for
the velocity gradient, which we found to be $127 \pm
25$\,km$^{-1}$\,s$^{-1}$\,arcsec$^{-1}$ along the slit. Assuming this
tilt, the [\ion{O}{2}] doublet is resolved, but the individual line
widths are consistent with the instrumental resolution. We measured a
total [\ion{O}{2}] flux from the host light in the slit of $(4.79 \pm
0.05) \times 10^{-17}$\,erg\,s$^{-1}$\,cm$^{-2}$, which is consistent
with the value previously reported \cite{chornock2013}.  With respect
to the [\ion{O}{2}] lines, we found that the \ion{Mg}{2} absorption
lines are blueshifted by $234 \pm 78 \pm 14$\,km\,s$^{-1}$, where the
first and second errors include only the uncertainty in the absorption
minima and only the uncertainty in the [\ion{O}{2}] maximum,
respectively. This is a higher value than previously reported, but,
based on our analysis, we derived a consistent value from the
published supernova spectrum ($202 \pm 13 \pm
107$\,km\,s$^{-1}$). This blueshift is larger than measured for most
star-forming galaxies observed at $z \sim 1.5$ \cite{erb2012}, but the
sample of such objects with blueshift measurements is small and shows
a large scatter (the outflow velocity measured for \afx's host is less
than a standard deviation larger than the sample mean).

For the foreground galaxy, the best fit emission line model prefers a
slight dependence on velocity with position, but this is not
significant. The total [\ion{O}{2}] flux from the foreground galaxy
light in the slit is $(1.48 \pm 0.08) \times
10^{-17}$\,erg\,cm$^{-2}$\,s$^{-1}$. From the initial, slightly offset
pointing the measured flux is about 50\% lower, which suggests the
East-West extent of the object is limited.

We note that [\ion{O}{2}] emission from the foreground galaxy is not
significantly detected in the published spectra of \afx\ obtained near
the supernova's peak brightness \cite{chornock2013}. If the emitting
region was fully contained in the slit aperture, a weak detection of
the [\ion{O}{2}] line could have been possible. This earlier spectrum
was, however, probably optimized for the extraction of the supernova
signal and may therefore only contain a fraction of the extended
foreground galaxy's light. Considering this, the lower signal-to-noise
ratio of the supernova spectrum, and the coincident location of the
[\ion{O}{2}] line with the blue edge of a broader supernova bump, it
is not surprising that the [\ion{O}{2}] line was not previously
identified.

\vspace{10mm}
\underline{Stellar Mass Estimates}
\vspace{5mm}

To test whether the foreground galaxy can satisfy the lensing
constraints for \afx, we must estimate its total mass. A galaxy's
rotation curve can be used to estimate a total mass \cite{rubin1980},
so we attempted to fit a 2-D model to the [\ion{O}{2}] emission line
in the observed spectrum (see section above). However, we do not detect a
significant dependence on the emission line's central wavelength as a
function of position. Doppler broadening of the absorption lines could
also indicate the foreground galaxy's mass
\cite{cappellari_emsellem2004}, but the data are confused with the
light of the more distant host galaxy, and the signal-to-noise ratio
available from the narrow strip outside of the host's glare is
prohibitively low. The \ion{Mg}{2} lines noted above are not resolved,
which suggests a 1-D velocity dispersion of less than about $\sigma <
90$\,km\,s$^{-1}$; yet, if they form in an outflow, the widths of
these lines may be decoupled from the galaxy's dynamical mass.

We instead use the foreground galaxy's stellar mass, which is
correlated with its 1-D velocity dispersion. We estimate stellar
masses for the galaxies by fitting the Bruzual \& Charlot (2003)
Single Stellar Population (SSP) models \cite{bruzual_charlot2003} to
our data. We choose models following the Padova 1994 evolutionary
tracks, and we tested both Salpeter and Chabrier stellar initial mass
functions (IMF). We assume our spectra are composed of two galaxies --
one at $z=1.3885$ and the other at $z=1.1168$ -- each with its own
internal extinction. The observed spectra are then modeled as a linear
combination of a subset of the models with ages between 0.1 and
5\,Gyr. Initially, we assumed each galaxy could be modeled by a bursty
star formation history and we allowed each to be comprised of SSP
models at four distinct ages all with 0.4 solar metallicity. However,
our fits suggest that each galaxy can be well represented by a single
age population. We then allowed each galaxy to have any single stellar
age in the $0.1 - 5$\,Gyr range and metallicities of 0.2, 0.4, or 1.0
solar. We find that young ages ($\sim$0.1\,Gyr) are strongly preferred
for the Host galaxy, and the foreground galaxy component is best fit
with ages close to 1\,Gyr.

To account for slit losses, we scaled our spectra by a factor of 1.4
to match the \zps-band magnitude reported for the ``host''
\cite{chornock2013}. As we have shown, this measurement must actually
be a combination of the light from the host and a foreground
galaxy. Thus, this scaling may not be perfectly valid for one or both
of the galaxies, but the significance of this effect should be small,
as noted below.

Assuming the Chabrier IMF, the best fit stellar masses are $(9 \pm 2)
\times 10^{9}$\,\msun\ for the foreground lens and $(7 \pm 1) \times
10^{9}$\,\msun\ for the more distant host. The metallicities for the
best fit models are 0.2 solar for the lens and 0.4 solar for the
host. These fits are slightly preferred to those with the Salpeter IMF,
which gives higher but consistent mass estimates: $(13 \pm 3) \times
10^{9}$\,\msun\ for the lens and $(10 \pm 1) \times
10^{9}$\,\msun\ for the host.

It should be noted that our spectra have relatively less flux below
7000\,\AA\ as compared to the ``host'' colors reported
\cite{chornock2013}. If we artificially de-redden our spectra assuming
E(B-V) = 0.3, we can recover these colors to within the
errors. However, these colors are not compatible with the measurements
from the CFHT Legacy Survey \cite{gwyn2008}, which, in particular,
favor significantly fainter g-band magnitudes for the host. It is thus
possible that the integrated color varies with the size and location
of the aperture, with the bluer light located outside of our slit
(although our two offset pointings show similar colors). Even with
this adjustment, however, the derived galaxy masses only change at the
$1-2\sigma$ level. If we, instead, normalize the photometry using the
reported g-band measurement \cite{chornock2013}, the galaxy masses
increase by less than a factor of 2. If the color discrepancy between
our spectra and the previous photometry is related to the fractions of
the light contributed to by the host and the lens, then this suggests
that any error in lens mass due to improperly scaling the light in the
slit to the total light measured from the photometry should also be
around a factor of 2 or less. Scaled to the g-band photometry, the
JHK magnitudes predicted by the best fit SSP models would, like the
redder optical bands, be well in excess of the observed
constraints. The predicted IR magnitudes are in the best agreement
with the observations (all within $1\sigma$) for the z-band scaled
case with no artificial reddening correction.

Higher masses for the foreground galaxy are possible if we assume an
older stellar population, but these produce worse fits to the
data. For example, if we assume a 5\,Gyr old population (close to the
maximum allowed for a $z=1.1168$ object), the mass increases to $50
\times 10^{9}$\msun, but the \ion{Ca}{2} K line, which lies in a
relatively clean spectral range, would be much stronger in the model
than allowed by the data. 

If we instead adopt a younger stellar population for the foreground
galaxy, then a large extinction is required. In this case, we would
expect the foreground galaxy to redden the light from \afx, but as
previously noted \cite{quimby2013b}, the observed colors suggest no
extinction of the supernova light. This problem could be avoided by
adopting an older age for the host's stellar population, but older SSP
models do not match the hydrogen Balmer absorption lines seen in the
break in the sky lines between 9000 and 9300\,\AA\ (as shown in
Fig. \ref{fig:ssp}, we clearly detect H9 and H10 from the host
including both broadened absorption dips from the stars and narrow
emission lines from gas).

The strongest stellar features expected from the lens galaxy are the
hydrogen Balmer and \ion{Ca}{2} H \& K lines. The signal-to-noise
ratio in this wavelength range is lower than in the range covering the
host features noted above, and the lens features are expected to be
weaker. As shown in figure \ref{fig:ssp_lens}, there is a possible
detection of H8, which lies in a relatively clean wavelength
range. The simple absorption dip predicted by the SSP modeling could
be complicated by nebular emission lines, which are not included in
the model. The data in the wings of the H8 feature appear consistent
with the SSP model, and there could be a narrow emission feature
emerging from the core of the line that would be consistent with
nebular light. A similar trend may hold for H$\delta$ and the
H$\epsilon$/\ion{Ca}{2} H blend, but interference from bright sky
lines precludes a definitive conclusion.

\vspace{10mm}
\underline{Lens Constraints}
\vspace{5mm}

We derive expected lens properties of \afx\ using a Monte Carlo
approach \cite{oguri_marshall2010}. We first convert our best stellar
mass estimate of $M_* = (9 \pm 2) \times 10^{9}$\,\msun\ for the
lensing galaxy to a velocity dispersion, $\sigma$. Using {\tt
  linmix\_err.pro} \cite{kelly2007}, we find the best fit linear
relation to the stellar masses and velocity dispersions of SDSS
galaxies \cite{kauffmann2003,blanton2005} is $\log{\sigma} = -1.4 +
0.33 \log{M_*}$, with an intrinsic scatter of 0.081 (see
Fig. \ref{fig:mstar-sigma}). Propagating the error from the stellar
mass estimate, we thus adopt a Gaussian distribution with a mean of
1.880 and dispersion of 0.088 for $\log(\sigma [{\rm km\,s^{-1}}])$
for the lensing galaxy. We assume the standard singular isothermal
ellipsoid mass distribution for the lens with an additional
contribution to the lens potential from external shear. The
ellipticity is Gaussian distributed with a mean of 0.25 and dispersion
of 0.2, and the magnitude of external shear follows a log-normal
distribution with a mean of 0.05 and dispersion of 0.2~dex. The
position angles are assumed to be random. Fixing the lens redshift to
$z=1.1168$, we randomly generate a list of lenses according to these
probability distributions of the velocity dispersion, ellipticity, and
external shear. For each lens, we uniformly distribute point sources
at $z=1.3885$, solve the lens equation using the public software, {\tt
  glafic} \cite{oguri2010}, and record any events that produce
multiple images. For all lenses we use the same angular number density
of sources, which suggests that the output catalog of multiple images
is automatically weighted by strong lensing cross sections.

With this procedure we generate mock catalog of over 500000 multiple
image sets. We then derive posterior distributions of any parameter
$X$ directly (e.g. the dashed red lines in Fig. \ref{fig:lens}) or,
for a consistency check with \afx, by adding a condition on the total
magnification $\mu$, $P(X)=\int P(X|\mu)P(\mu) d\mu$ (the blue
histograms in Fig. \ref{fig:lens}), where $P(\mu)$ is assumed to be
Gaussian distribution with a mean of 31 and dispersion of 5, which
corresponds to the best-fit and error of the total magnification of
\afx\ \cite{quimby2013b}. When so weighted by the lensing probability,
the velocity dispersion for the lensing galaxy is found to be
$\log(\sigma) = 1.95 \pm 0.09$, the image separation is $\log(\Delta
\theta) = -1.20 \pm 0.18$ (with $\Delta \theta$ in arcseconds), and
the time delay is $\log(\Delta t) = -0.61 \pm 0.44$ (with $\Delta t$
in days). We also find that the centroid of the combined supernova
images should be offset from the lens center by $\log( \theta) = -1.68
\pm 0.26$ (with $\theta$ in arcseconds).

These values are fully consistent with the observations of \afx. To be
inconsistent, the velocity dispersion would have to be significantly
larger than we have estimated. If we adopt the velocity dispersion
implied by the highest stellar mass considered above ($50 \times
10^{9}$\msun\ or $\log(\sigma) = 2.13$), for example, the expected
lens parameters would still be consistent with the observations of
\afx. This strengthens our conclusion that the foreground galaxy is
the gravitational lens that has magnified \afx.

We perform an additional plausibility check that is completely
independent of our stellar mass estimate. We consider the probability
distribution of the 1-D velocity dispersion of galaxies at $z=1.1168$
that can act as a lens for a $z=1.3883$ source. For a spherically
symmetric gravitational lens with an isothermal density profile
characterized by a velocity dispersion $\sigma$, the Einstein radius
is given by,
\begin{equation}
b=4\pi\,\left(\frac{\sigma}{c}\right)^2 \frac{D_{\rm ls}}{D_{\rm s}}\,,
\label{einst_rad_sis}
\vspace{0.1cm}
\end{equation}
\noindent where $D_{\rm ls}$ and $D_{\rm s}$ are angular diameter
distances from lens to the source and from observer to the source,
respectively. The lens cross-section for a spherical isothermal model
is given by ${\cal A}_{\rm lens}=\pi b^2 \propto \sigma^4$. Therefore,
the probability distribution of the velocity dispersion of a galaxy at
redshift $z_{\rm l}$ to lens a background galaxy at redshift $z_{\rm
  s}$ is given by
\begin{equation}
P(\sigma|z_{\rm l},z_{\rm s}) \propto n(\sigma,z_{\rm l}) {\cal A}_{\rm lens} \propto
n(\sigma,z_{\rm l})\sigma^4\,.
\label{psigma}
\end{equation}

We use the inferred velocity dispersion function for the sample of
galaxies in the redshift range $0.9\le z \le 1.2$
\cite{bezanson2011}. We fit a Schechter function to these data points,
and extrapolate the resulting velocity dispersion function to lower
values of the velocity dispersion. The probability distribution
function for the lens velocity dispersion is then obtained by
inserting this velocity dispersion function into
Equation~\ref{psigma}. We have restricted ourselves to those velocity
dispersions which result in image separation distribution less than
$0.4''$, given that \afx\ was not resolved by Pan-STARRS imaging
data. The angular size of the SN photosphere must be smaller than the
SN$-$lens angular offset, $\Delta \theta$, to avoid changes in the
effective magnification over time. This puts a constraint on the lower
mass end of the velocity dispersion. After the application of these
priors, the median velocity dispersion,
$\sigma=177^{+32}_{-47}\,$\,\kps, is compatible if not slightly higher
than the value derived through our Monte Carlo simulation. This value
is also consistent (albeit at the higher end) of the $\sigma - M_*$
relation derived from the SDSS galaxies, which is not entirely
unexpected given that lensing probability grows rapidly as $\sigma$
increases (Eqn.~\ref{psigma}).

\vspace{10mm}
\underline{Selecting Lensed \SNIa\ in Color-Magnitude Space}
\vspace{5mm}

To determine the distribution of lensed and un-lensed supernovae in
color-magnitude space, we first calculate the observed magnitudes
expected in the $r$ and $i$-bands for \SNIa\ and core-collapse
supernovae as a function of redshift. For band $X$, the observed
magnitude is computed from $m_X = M + \mu + K_X$, where $\mu$ is the
distance modulus, $M$ is the absolute magnitude of the supernova in
some rest frame band, and the $K_X$ term accounts for the offset
between this band and band $X$ including the effects of redshift. We
allow for a range of peak absolute magnitudes for each type of
supernova \cite{li2011a}. We calculate the so-called $K$-corrections
using spectral energy distribution templates for various supernovae
\cite{nugent_templates,hsiao2007}. We use the rates of supernovae
\cite{graur2011} to determine the relative numbers of each type of
supernovae with redshift.

The sample of supernovae generated from this is then used as input to
a Monte Carlo simulation that determines what fraction of events are
lensed and by how much. For the lensing galaxies, we adopt the
velocity function measured from SDSS galaxies \cite{bernardi2010}. We
assume no redshift evolution. The strong lensing probability is
calculated assuming a singular isothermal ellipsoid with a fixed
ellipticity of 0.3. For each lensing event, the total magnification is
computed using {\tt glafic} \cite{oguri2010}. With the updated
velocity function and ellipticity, we find the lensing probability is
a factor of 2 higher than previously published
\cite{oguri_marshall2010}. Our simulations show that for a flux
limited survey, the $r-i$ colors of most lensed supernovae can be
significantly larger (redder) than the un-lensed events. Objects found
to lie above the bold line in figure \ref{fig:cmd} can thus be
considered lensed supernova candidates.

We can verify the purity of such selection criteria using the
published sample of supernova discoveries from the Supernova Legacy
Survey (SNLS) \cite{guy2010}. Unfortunately, only the set of well
characterized events (e.g. \SNIa) is available but we can at least use
these to verify that the dominant product of similar supernovae
surveys will be excluded from our selection cuts. The filters used in
the SNLS Survey differ slightly from the standard SDSS band passes,
which leads to slightly different selection criteria than depicted in
figure \ref{fig:cmd}. Accounting for this, we compare the observed
$i$-band magnitudes and $r-i$ colors of the SNLS sample and find that
all events detected before maximum light have photometry that lies
below our selection line at some point before declining. At late
times, the colors can be redder, but selecting only those objects that
are consistently above our (appropriately modified) selection line
during their rise to maximum light eliminates the entire SNLS
sample. The selection criteria thus succeeds in weeding out normal
supernovae during their rising phase and enables the selection of
lensed \SNIa\ on the rise, which will enable good constraints on the
date of and flux at maximum light.

\vspace{10mm}
\underline{Selection Bias and the Expected Number of Lensed \SNIa\ from Pan-STARRS1}
\vspace{5mm}

Even though the magnification of \afx\ is larger than expected from a
random, strong-lensing system in a given volume and the lens galaxy is
less massive than a random draw weighted by lensing cross section
would predict, the system may well prove typical of the
gravitationally lensed \SNIa\ discoveries that will be found by future
surveys. Flux limited surveys like Pan-STARRS1 have a strong bias
against less magnified \SNIa, which are fainter thus harder to
detect. Because both lensing optical depth and the number of \SNIa\ is
a steep function of redshift (especially around $z\sim1$) due to
increasing comoving volume and, in the case of \SNIa, rates, there
are more strongly lensed \SNIa\ at higher redshifts with large
magnifications than at lower redshifts with lower magnifications (and
thus similar observed magnitudes). There is thus a strong overall bias
for highly magnified \SNIa. Also, the angular resolution of
Pan-STARRS1 is sufficient to resolve lensed \SNIa\ with large image
separations, which are necessarily seen through the most massive lens
galaxies, but the SN images seen through smaller lens galaxies are
more likely to be unresolved. Small lens galaxies thus have a
selection advantage in flux limited searches as the multiple supernova
images will blend together to form a single point source with a flux
equal to the sum of its parts; in the case of resolved images, the
individual images must be detected, but these will of course be
fainter.

We illustrate this point in figure \ref{fig:sepdist}, which shows the
distribution of gravitationally lensed \SNIa\ expected to be detected
by the Pan-STARRS1 Medium Deep Survey (PS1-MDS). We first used a Monte
Carlo simulation similar to one previously published (hereafter OM10)
\cite{oguri_marshall2010} to predict the distribution that would be
found using prior selection techniques. Specifically, detection of the
third brightest image was required for quads and detection of the
second image was required for doubles. Excluding unresolved systems,
the total number of gravitationally lensed \SNIa\ predicted by this
method is 0.1 over the survey lifetime, which is consistent with the
number published in OM10. However, if we only require detection of a
single image (or a single blend in the case of unresolved images), the
expected number of gravitationally lensed \SNIa\ from the PS1-MDS
jumps to 0.9. About half of these should be unresolved by the survey,
like \afx. The Monte Carlo simulation further predicts that a
\SNIa\ sample selected in this way will have a mean total
magnification of 13.0, a median of 5.0, and 95\% of the sample will
have total magnifications in the range $2.0 < \mu < 59.2$. This,
again, is consistent with \afx.

\renewcommand{\thefigure}{S1} 
\begin{figure}
\begin{center}
 \includegraphics[width=\linewidth]{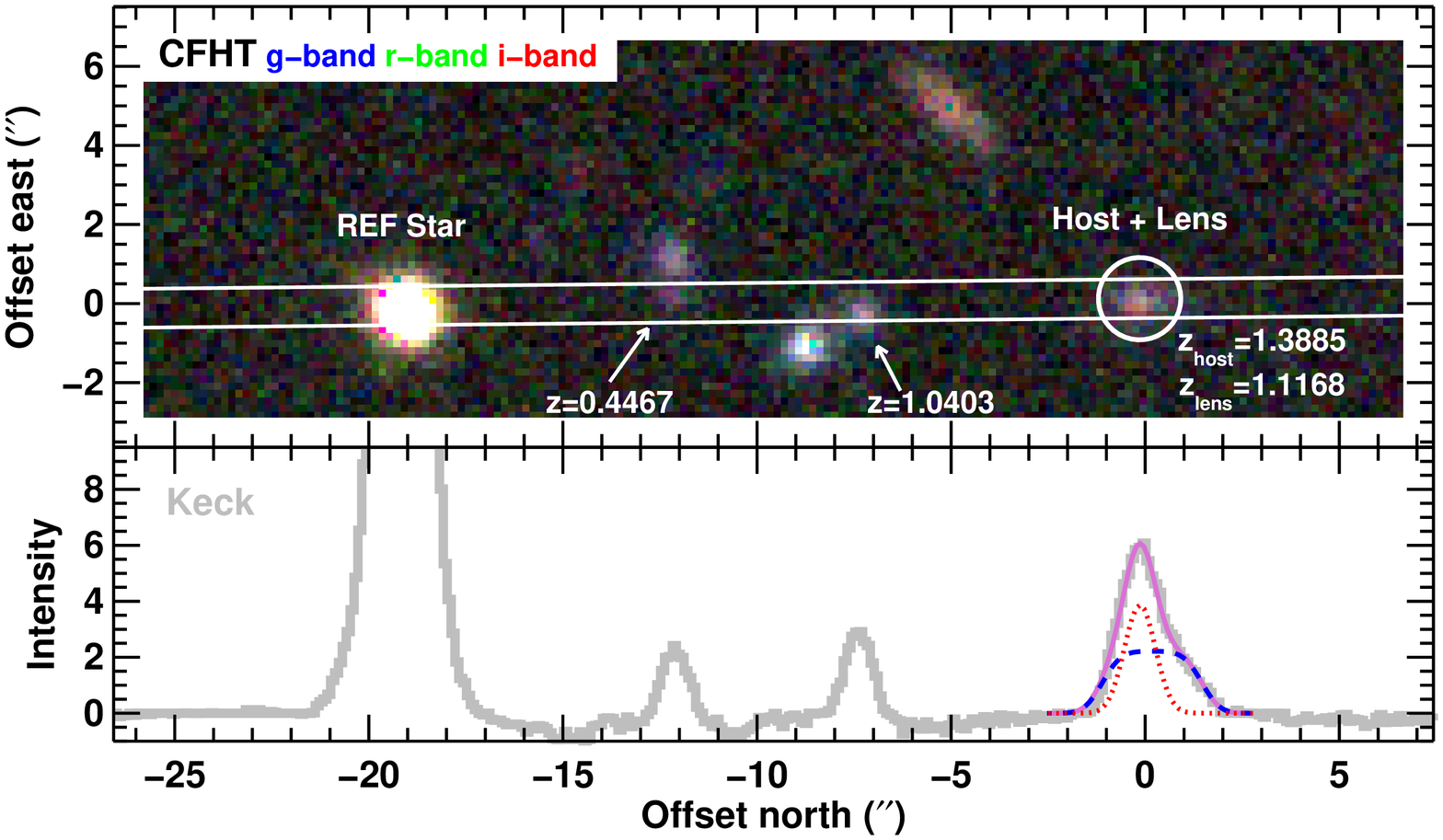} 
 \caption{ Field setup for the Keck/LRIS observation of \afx. The top
   panel shows a color-composite image of the sky near \afx\ using g,
   r, and i-band data taken prior to the outburst by the
   Canada-France-Hawaii Telescope Legacy Survey \cite{gwyn2008}. The
   lines mark the location of the 1.0\arcsec\ slit mask deployed for
   spectroscopy. The location of \afx\ is marked with a white circle,
   and the redshifts of nearby galaxies, as determined from the Keck
   spectra, are indicated. The lower panel shows the 1-D intensity
   along the slit as recorded by the Keck observation. The target
   profile (purple line) was decomposed into a marginally resolved
   host component (red dotted line) and an extended foreground galaxy
   component (blue dashed line).}
   \label{fig:slit}
\end{center}
\end{figure}

\renewcommand{\thefigure}{S2} 
\begin{figure}
\begin{center}
 \includegraphics[width=\linewidth]{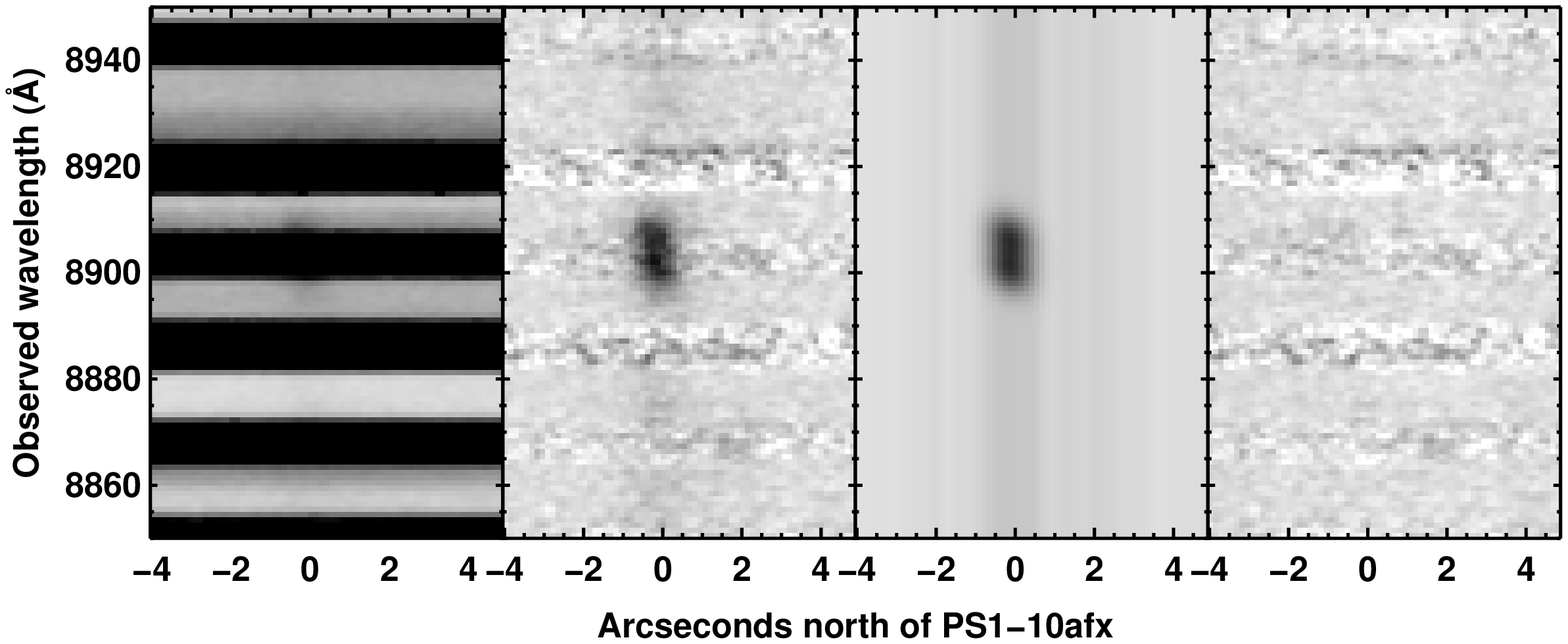} 
 \caption{ Composite 2-D spectra showing the [\ion{O}{2}] emission
   from the host at $z=1.3885$. The first two panels on the left show
   the stacked spectra before and after removal of the sky
   background. The next panel shows the 2-D model for the emission
   lines and continuum, and the right panel shows the residual after
   this model was subtracted from the second panel. }
   \label{fig:host_OII}
\end{center}
\end{figure}

\renewcommand{\thefigure}{S3} 
\begin{figure}
\begin{center}
 \includegraphics[width=\linewidth]{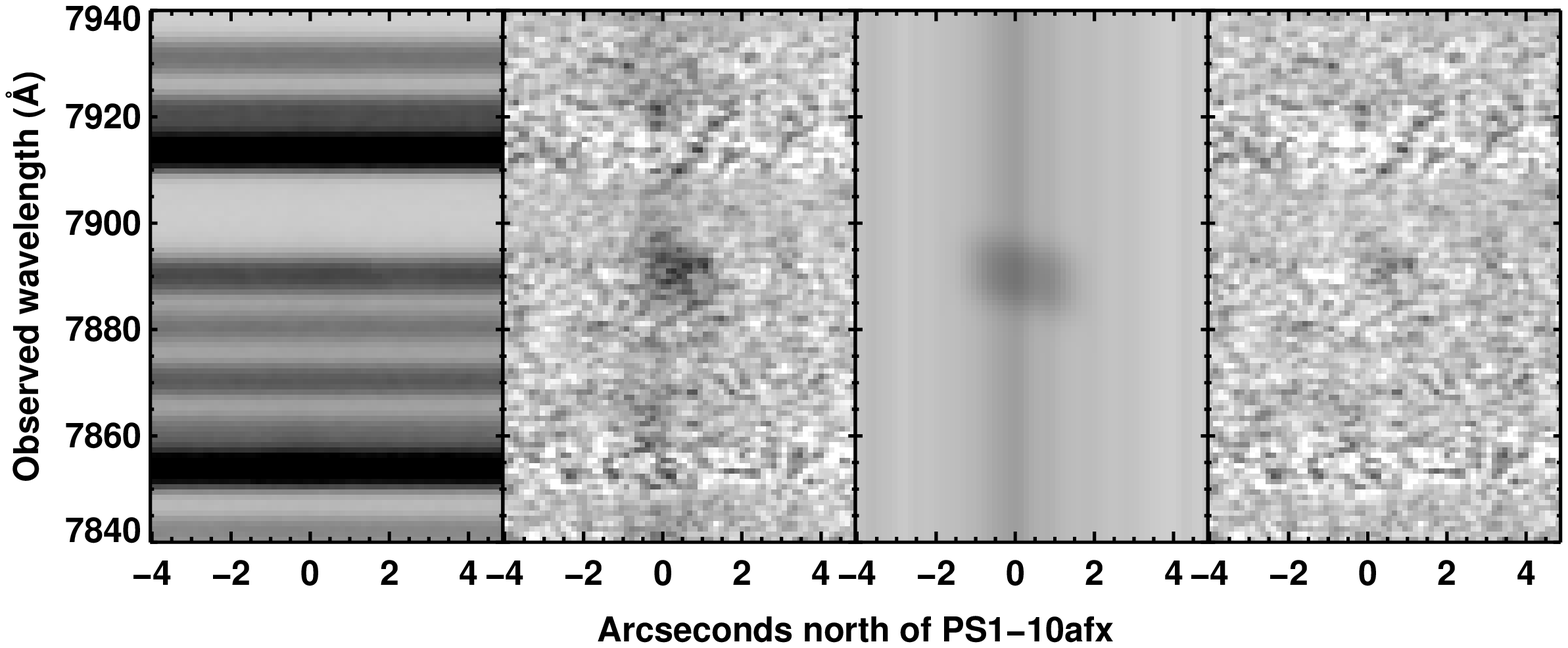} 
 \caption{ Similar to figure \ref{fig:host_OII} but showing the
   [\ion{O}{2}] emission from the foreground galaxy at $z=1.1168$. }
   \label{fig:lens_OII}
\end{center}
\end{figure}

\renewcommand{\thefigure}{S4} 
\begin{figure}
\begin{center}
 \includegraphics[width=\linewidth]{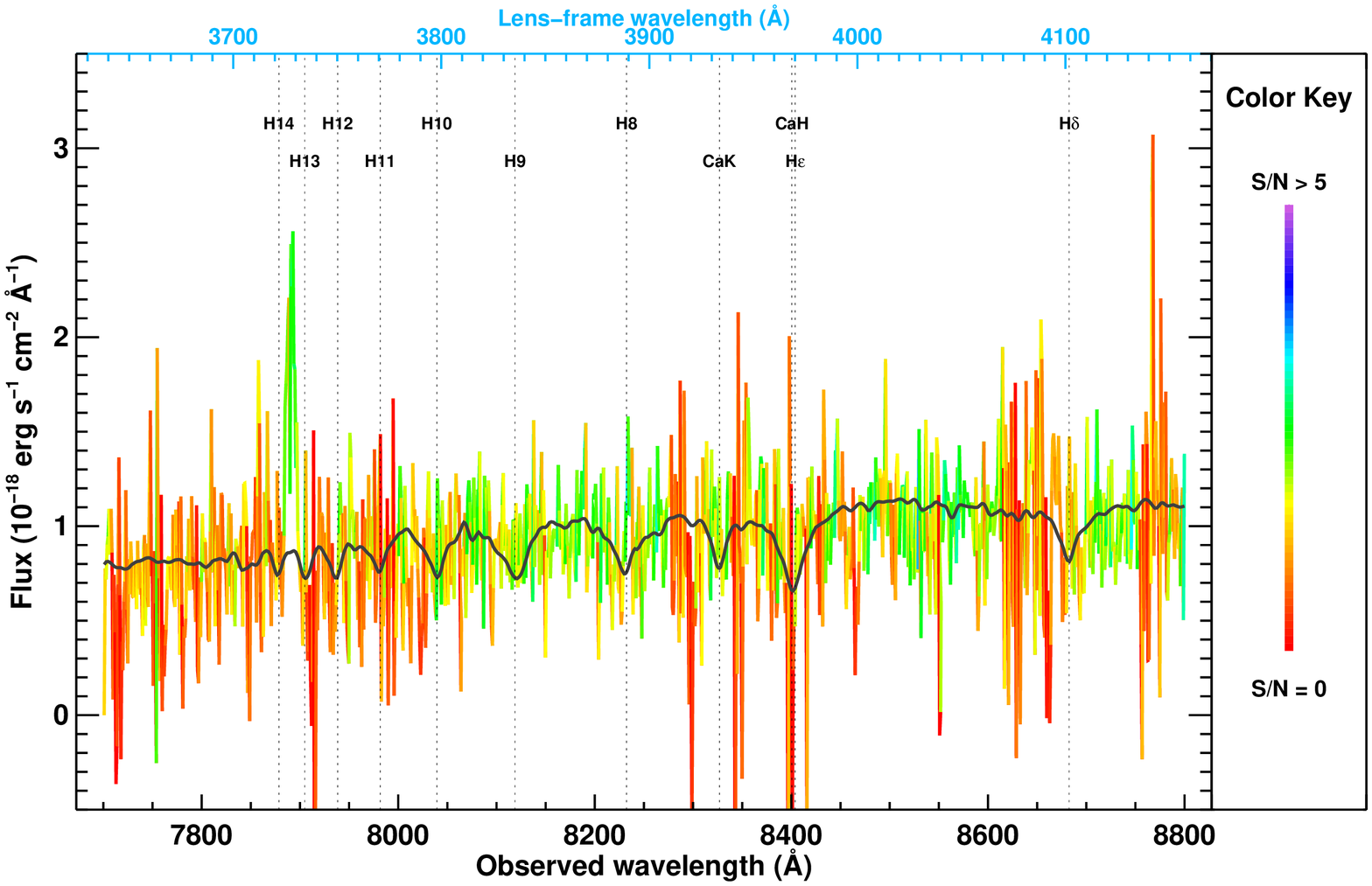} 
 \caption{Detail of the best fit SSP model showing the strongest
   stellar lines expected from the lens galaxy. The data are color
   coded by the signal-to-noise ratio (per Angstrom) as indicated in
   the legend. The dark gray curve is the best fit SSP model, which
   includes the total starlight expected from the host and lens but
   neglects emission from nebular gas.}
   \label{fig:ssp_lens}
\end{center}
\end{figure}

\renewcommand{\thefigure}{S5} 
\begin{figure}
\begin{center}
 \includegraphics[width=\linewidth]{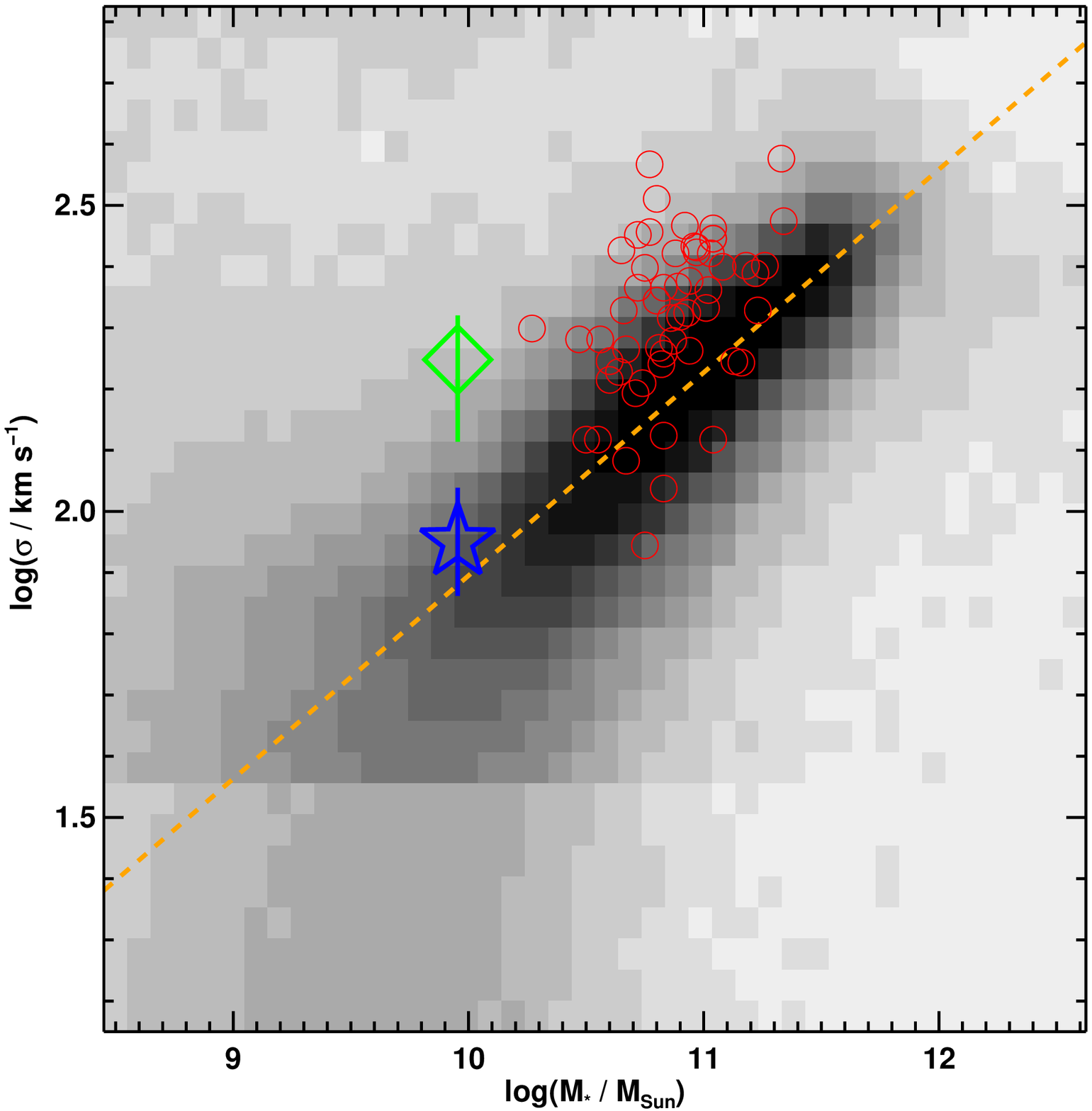} 
 \caption{ Stellar mass ($M_*$) vs. velocity dispersion ($\sigma$)
   relation for galaxies. The gray scale contours show the density of
   galaxies from the SDSS \cite{kauffmann2003,blanton2005} on a cube
   root scale. The dashed orange line shows the best fit linear
   relation. The blue star marks the most probable $\sigma$ for the
   lensing galaxy including the magnification prior from our Monte
   Carlo simulation, and the green diamond is the estimated value from
   the analytic calculation. The red circles are for a sample of
   galaxies at redshifts comparable to the lens \cite{belli2013}.}
   \label{fig:mstar-sigma}
\end{center}
\end{figure}

\renewcommand{\thefigure}{S6} 
\begin{figure}
\begin{center}
 \includegraphics[width=\linewidth]{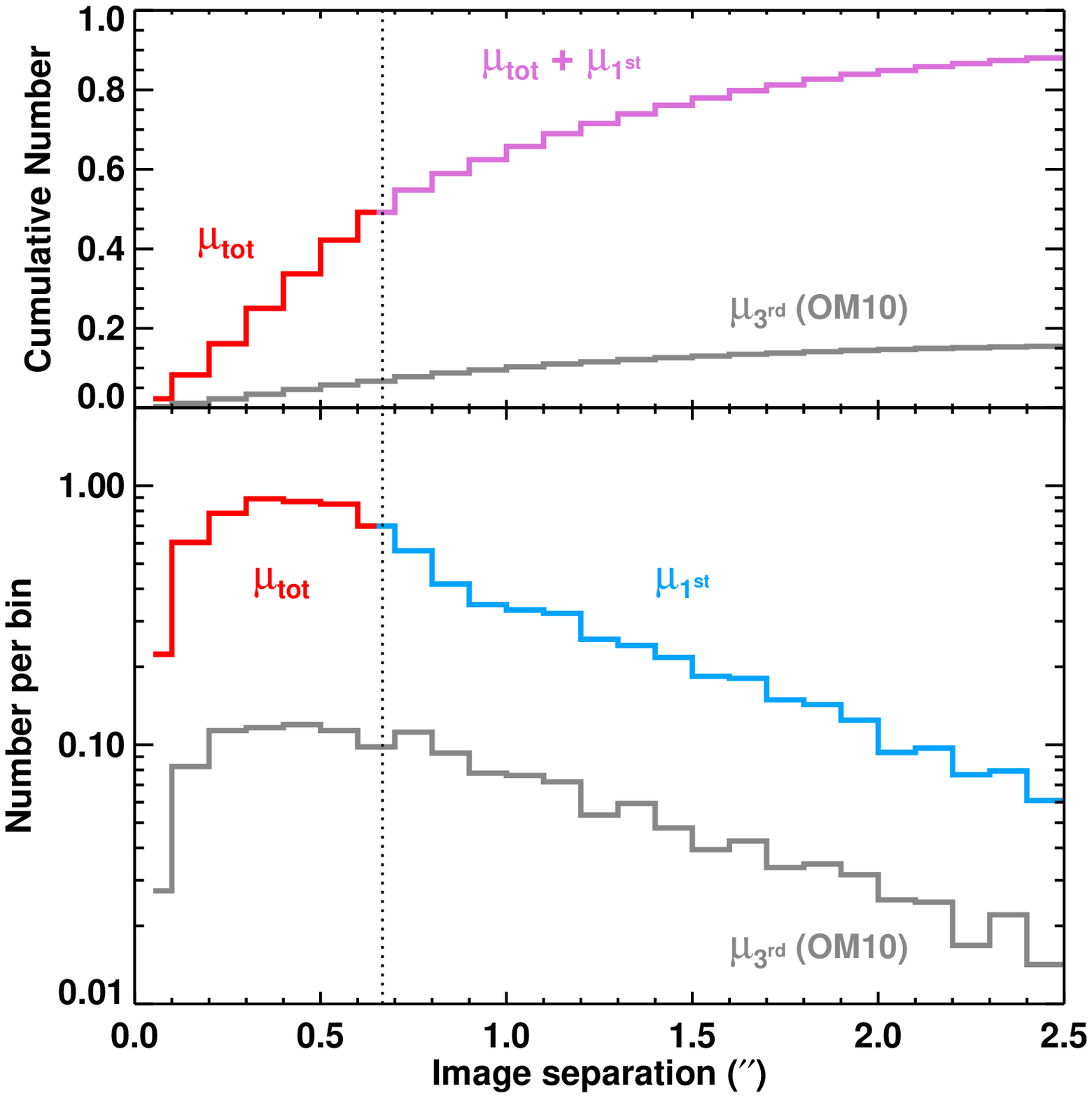} 
 \caption{Expected number distribution of lensed \SNIa\ from the
   PS1-MDS as a function of image separation angle (the distance
   between the two images in a double or the maximum separation in a
   quad system). Predictions from OM10 \cite{oguri_marshall2010} were
   based on the detectability of the fainter image in a double and the
   third brightest image in a quad (gray lines in the figure). Using
   the color selection, lensed supernovae can then be identified
   solely from the brightest image in the system ($\mu_{1^{\rm st}}$;
   blue line). When the image separation is below the resolving power
   of the survey (vertical dotted line), the relevant brightness is
   set by the sum of the individual images ($\mu_{tot}$; red
   lines). The total number of lensed \SNIa\ detectable by PS1-MDS is
   then roughly an order of magnitude larger than previously
   considered.}
   \label{fig:sepdist}
\end{center}
\end{figure}


\begin{thebibliography}{10}

\bibitem{chornock2013}
R.~{Chornock}, {\it et~al.\/}, {\it \apj\/} {\bf 767}, 162 (2013).

\bibitem{galyam2012}
A.~{Gal-Yam}, {\it Science\/} {\bf 337}, 927 (2012).

\bibitem{rakavy1967}
G.~{Rakavy}, G.~{Shaviv}, Z.~{Zinamon}, {\it \apj\/} {\bf 150}, 131 (1967).

\bibitem{barkat1967}
Z.~{Barkat}, G.~{Rakavy}, N.~{Sack}, {\it Physical Review Letters\/} {\bf 18},
  379 (1967).

\bibitem{woosley2007}
S.~E. {Woosley}, S.~{Blinnikov}, A.~{Heger}, {\it \nat\/} {\bf 450}, 390
  (2007).

\bibitem{woosley2010}
S.~E. {Woosley}, {\it \apjl\/} {\bf 719}, L204 (2010).

\bibitem{kasen_bildsten2010}
D.~{Kasen}, L.~{Bildsten}, {\it \apj\/} {\bf 717}, 245 (2010).

\bibitem{chevalier_irwin2011}
R.~A. {Chevalier}, C.~M. {Irwin}, {\it \apjl\/} {\bf 729}, L6 (2011).

\bibitem{moriya2013}
T.~J. {Moriya}, {\it et~al.\/}, {\it \mnras\/} {\bf 428}, 1020 (2013).

\bibitem{quimby2013b}
R.~M. {Quimby}, {\it et~al.\/}, {\it \apjl\/} {\bf 768}, L20 (2013).

\bibitem{phillips1999}
M.~M. {Phillips}, {\it et~al.\/}, {\it \aj\/} {\bf 118}, 1766 (1999).

\bibitem{jha2007}
S.~{Jha}, A.~G. {Riess}, R.~P. {Kirshner}, {\it \apj\/} {\bf 659}, 122 (2007).

\bibitem{hicken2009}
M.~{Hicken}, {\it et~al.\/}, {\it \apj\/} {\bf 700}, 1097 (2009).

\bibitem{sullivan2010}
M.~{Sullivan}, {\it et~al.\/}, {\it \mnras\/} {\bf 406}, 782 (2010).

\bibitem{oguri_marshall2010}
M.~{Oguri}, P.~J. {Marshall}, {\it \mnras\/} {\bf 405}, 2579 (2010).

\bibitem{som}
Additional information is available in the supplementary materials.

\bibitem{oke1995}
J.~B. {Oke}, {\it et~al.\/}, {\it \pasp\/} {\bf 107}, 375 (1995).

\bibitem{rockosi2010}
C.~{Rockosi}, {\it et~al.\/}, {\it Society of Photo-Optical Instrumentation
  Engineers (SPIE) Conference Series\/} (2010), vol. 7735 of {\it Society of
  Photo-Optical Instrumentation Engineers (SPIE) Conference Series\/}.

\bibitem{erb2012}
D.~K. {Erb}, A.~M. {Quider}, A.~L. {Henry}, C.~L. {Martin}, {\it \apj\/} {\bf
  759}, 26 (2012).

\bibitem{kauffmann2003}
G.~{Kauffmann}, {\it et~al.\/}, {\it \mnras\/} {\bf 341}, 33 (2003).

\bibitem{blanton2005}
M.~R. {Blanton}, {\it et~al.\/}, {\it \aj\/} {\bf 129}, 2562 (2005).

\bibitem{refsdal1964}
S.~{Refsdal}, {\it \mnras\/} {\bf 128}, 307 (1964).

\bibitem{oguri_kawano2003}
M.~{Oguri}, Y.~{Kawano}, {\it \mnras\/} {\bf 338}, L25 (2003).

\bibitem{filippenko1982}
A.~V. {Filippenko}, {\it \pasp\/} {\bf 94}, 715 (1982).

\bibitem{hanuschik2003}
R.~W. {Hanuschik}, {\it \aap\/} {\bf 407}, 1157 (2003).

\bibitem{iraf}
IRAF is distributed by the National Optical Astronomy Observatory, which is
  operated by the Association of Universities for Research in Astronomy (AURA)
  under cooperative agreement with the National Science Foundation.

\bibitem{buton2013}
C.~{Buton}, {\it et~al.\/}, {\it \aap\/} {\bf 549}, A8 (2013).

\bibitem{van-dokkum2001}
P.~G. {van Dokkum}, {\it \pasp\/} {\bf 113}, 1420 (2001).

\bibitem{kelson2003}
D.~D. {Kelson}, {\it \pasp\/} {\bf 115}, 688 (2003).

\bibitem{rubin1980}
V.~C. {Rubin}, W.~K.~J. {Ford}, N.~{.~Thonnard}, {\it \apj\/} {\bf 238}, 471
  (1980).

\bibitem{cappellari_emsellem2004}
M.~{Cappellari}, E.~{Emsellem}, {\it \pasp\/} {\bf 116}, 138 (2004).

\bibitem{bruzual_charlot2003}
G.~{Bruzual}, S.~{Charlot}, {\it \mnras\/} {\bf 344}, 1000 (2003).

\bibitem{gwyn2008}
S.~D.~J. {Gwyn}, {\it \pasp\/} {\bf 120}, 212 (2008).

\bibitem{kelly2007}
B.~C. {Kelly}, {\it \apj\/} {\bf 665}, 1489 (2007).

\bibitem{oguri2010}
M.~{Oguri}, {\it \pasj\/} {\bf 62}, 1017 (2010).

\bibitem{bezanson2011}
R.~{Bezanson}, {\it et~al.\/}, {\it \apjl\/} {\bf 737}, L31 (2011).

\bibitem{li2011a}
W.~{Li}, {\it et~al.\/}, {\it \mnras\/} {\bf 412}, 1441 (2011).

\bibitem{nugent_templates}
See http://supernova.lbl.gov/$\sim$nugent/nugent\_templates.html.

\bibitem{hsiao2007}
E.~Y. {Hsiao}, {\it et~al.\/}, {\it \apj\/} {\bf 663}, 1187 (2007).

\bibitem{graur2011}
O.~{Graur}, {\it et~al.\/}, {\it \mnras\/} {\bf 417}, 916 (2011).

\bibitem{bernardi2010}
M.~{Bernardi}, {\it et~al.\/}, {\it \mnras\/} {\bf 404}, 2087 (2010).

\bibitem{guy2010}
J.~{Guy}, {\it et~al.\/}, {\it \aap\/} {\bf 523}, A7 (2010).

\bibitem{belli2013}
S.~{Belli}, A.~B. {Newman}, R.~S. {Ellis}, {\it ArXiv e-prints\/}  (2013).

\end{thebibliography}
\end{document}